\documentstyle[preprint,eqsecnum,aps]{revtex}

\begin{document}
\draft
\preprint{HEP/123-qed}
\title{Polaron Self-trapping in a Honeycomb Net}
\author{Marco Zoli}
\address{Istituto Nazionale di Fisica della Materia - Universit\'a di Camerino, \\
62032 Camerino, Italy. e-mail: zoli@campus.unicam.it
}

\date{\today}
\maketitle
\begin{abstract}
Small polaron behavior in a two dimensional
honeycomb net is studied by applying the
strong coupling perturbative method to the
Holstein molecular crystal model.
We find that small optical polarons can be
mobile also if the electrons are strongly
coupled to the lattice. Before the polarons
localize and become very heavy, there is infact a window of {\it e-ph} couplings in which the
polarons are small and have masses of order
$\simeq 5 - 50$ times the bare band mass
according to the value of the adiabaticity
parameter. The 2D honeycomb net favors the
mobility of small optical
polarons in comparison with the square lattice.
\end{abstract}
\narrowtext
%\widetext
\section*{I.Introduction}

The transport properties of real systems
are strongly affected by the presence of
non linear potentials in the lattice
\cite{nfmott,ssh,devreese}.
Nonlinearities may arise because of embedded impurities 
in the host lattice
which favor trapping of the charge carriers
\cite{wipf,scharf,molina,gupta} in
any dimensionality \cite{dunlap} or, nonlinearities may be intrinsic to the system and driven by the
electron-lattice interaction as assumed in
the Molecular Crystal Model first proposed
by Holstein \cite{holst59} in the form of a discrete 
nonlinear Schr\"odinger equation for electrons 
coupled to harmonic phonons.
Several theoretical methods have been
developed in the last decades 
\cite{emi,raedt,kopida,pucci,jeckel,mello,robin,ale,rome}
to analyse the ground state and finite
temperature properties of the unit comprising the electron plus the surrounding 
local lattice deformation, namely the polaron.
Central to these investigations is the concept
of self-trapping traditionally denoting a
transition between an infinite radius state
at weak electron-phonon coupling and a finite radius polaron at strong e-ph coupling. The narrowing of the polaron band
and the abrupt increase of the polaron effective mass versus e-ph coupling are the classical and related indicators 
of the transition event which may 
occur or not according to the degree of
adiabaticity and the dimensionality of the
system.
When the characteristic phonon energy $\bar \omega$
becomes larger than the electronic bandwidth
the antiadiabatic regime is attained. In this case 
it is  generally accepted that the
polaron wave function spreads over a few
lattice sites (small polaron) with the
polaron mass being a smooth function of the
e-ph coupling. 
Instead, the polaron self-traps in the adiabatic regime 
and there is growing evidence
that this event takes place in any dimensionality \cite{jeckel,rome,io}.

When the lattice polarizations is confined
to one or a few unit cells the carrier feels
the details of the local structure:
recent generalizations of the Holstein model
have shown that the inclusion of on site lattice 
anharmonicity can substantially
modify the size \cite{chris} and the mass 
\cite{voulga} of the polaronic quasiparticle.
While these findings could contribute to locate
with more accuracy the self-trapping event in parameter
space it is still unclear whether and how the 
transition depends on the lattice structure.
To address this problem we focus here
on the polaron mass 
renormalization in a two dimensional honeycomb net
which can be viewed as a triangular Bravais lattice
with a bases of diatomic molecules at the vertexes.
The Holstein model is briefly reviewed 
and the results are discussed in Section II.
The conclusions are drawn in Section III.

\section*{II. The Holstein Model Hamiltonian}

The  Hamiltonian for the single electron in the 
Holstein model reads

\begin{equation}
H= -t \sum_{i \ne j }c_i^{\dag}c_{j}
+ g \sum_i c_i^{\dag}c_i (a_i + a_i^{\dag})
+ \sum_{\bf k} \omega_{\bf k}a^{\dag}_{\bf k}a_{\bf k}
\label{1}
\end{equation}

where the  dimension dependence explicitly appears
in the momentum space Hamiltonian for the harmonic
lattice vibrations. $c_i^{\dag}$ ( $c_{i}$ ) creates
(destroys) a tight binding electron at the $i$ site
and $t$ is the first neighbor hopping integral related to the
bare electron half bandwidth $D$ by $D=zt$, $z$ being
the coordination number. 
$a_{\bf k}^{\dag}$ ( $a_{\bf k}$ ) creates
(destroys) a {\bf k}-phonon with frequency 
$\omega_{\bf k}$. 
$g$ is the overall electron-phonon coupling constant. 

In the strong coupling regime the Lang-Firsov approach \cite{lang} is
reliable \cite{gogo,firsov} and the polaron mass $m^*$
can be obtained via perturbative method. 
In $d$ dimensions
the ratio between $m^*$ and the bare band mass $m_0$ is \cite{io}:

\begin{eqnarray}
\Bigl( {{m^*} \over {m_0}}\Bigr)_d=\,& &
{{ exp(\bar g^2)} \over {1 + z^2t 
exp(- \bar g^2)
f(\bar g^2)/(\hbar \bar \omega)}} \,
\nonumber \\
\bar g^2=\,& & 
{{2g^2} \over {N}}\sum_{k_x} \sin^2 {{k_x } \over 2}
\sum_{k_x, k_y} (\hbar \omega_{\bf k})^{-2} \,
\nonumber \\
f(\bar g^2)=\,& & \sum_{s=1}^{+ \infty}
{{(\bar g^2)^s} \over {s s!}}
\label{2}
\end{eqnarray}

The series expansion in the last of eqs.(2) reflects
the fact that the second order polaron self energy 
comprises the emission and absorption of an arbitrary
number of phonons hence it is a sum over
an infinite number of diagrams each having $s$ phonons
between the two interaction vertexes. 
The second order of
perturbative theory also introduces the effect of the
adiabaticity parameter $zt/\hbar \bar \omega$ on $m^*$ which, in general,
depends on dimensionality through: 
i) $g^2 \propto d$, 
ii) the first neighbors number $z$, iii) the
Brillouin zone sums and
iv) the features of the phonon spectrum.

We take a 2D honeycomb net equivalent to a triangular
lattice with a two points basis. Each lattice site
is a diatomic molecule with coordination number $z=\,6$.
Hence, the phonon spectrum has both acoustical and optical
branches whose analytical expressions can be deduced
by a force constants parametrization scheme:

\begin{eqnarray}
\omega^2(k_x,k_y)=\,& & {{\beta + 3\gamma}\over M} 
\pm {1 \over M} \cdot 
\sqrt{\beta^2 + \gamma^2H(k_x,k_y) + \beta \gamma G(k_x,k_y)} \,
\nonumber \\
H(k_x,k_y)=\,& & 3 + 2(c_xc_y + s_xs_y + c_{3x}c_y + 
s_{3x}s_y + c_{2x}) \,
\nonumber \\
G(k_x,k_y)=\,& & 2(2c_xc_y + c_{2x})
\label{3}
\end{eqnarray}

with: $c_x=\, cos(k_x \sqrt{3}a/2)$, $c_y=\, cos(k_y 3a/2)$,
$c_{2x}=\, cos(k_x \sqrt{3}a)$, 
$c_{3x}=\, cos(k_x 3\sqrt{3}a/2)$,
$s_x=\, sin(k_x \sqrt{3}a/2)$, $s_y=\, sin(k_y 3a/2)$,
$s_{3x}=\, sin(k_x 3\sqrt{3}a/2)$. $a$ is the lattice constant and $M$ is the reduced molecular mass.
$\beta$ and $\gamma$ are the {\it intra}- and {\it inter}-molecular force constants respectively in terms
of which one defines: $\omega_0^2=\, 2\beta/M$, 
$\omega_1^2=\, \gamma/M$ and the zone center optical
frequency $\bar \omega=\, \sqrt{\omega_0^2 + z \omega_1^2}$.

Previous investigations of the Holstein Hamiltonian
\cite{io1} have shown that the intermolecular forces
have to be sufficiently strong in order to predict
the correct polaron bandwidth trend versus
dimensionality. Thus the ground state properties of
the Holstein model essentially depend on the strength of $\omega_1$ which should be of
order $\simeq \omega_0/2$. When this condition is
fulfilled the polaron mass turns out to be substantially
dimension independent. Larger
$\omega_1$ values are admitted in the model although they encounter the obvious
upper bound $\omega_1 \le \omega_0$ in a molecular lattice. This result (which
has been proven  in a large portion of
parameter space ranging from fully adiabatic to
antiadiabatic conditions) introduces
a novel feature in the polaron landscape corroborated 
by Monte Carlo simulations \cite{kornilo}, density matrix
renormalization-group studies \cite{jeckel} 
and variational approaches. 

Two quantities play a central role in polaron theory.
The first, defined by 
$$\lambda=\, {{N g^2}\over 
{D \sum_{\bf k} \hbar \omega_{\bf k}}}$$
$N$ being the number of molecular sites, represents
the ratio \cite{io2}
between the polaron binding energy and the
electronic half bandwidth. It yields the energetical
gain due to small polaron formation 
with respect to the bare electronic state.
The second, defined by 
$$\alpha=\, {{N g}\over 
{\sum_{\bf k} \hbar \omega_{\bf k}}}$$,
measures the lattice deformation associated with the quasiparticle formation.
While in adiabatic systems the condition
$\lambda > 1$ signals the existence of the small
polaron state, in antiadiabatic systems 
$\alpha > 1$
is a more restrictive condition for small polaron formation \cite{io3}.

Recent analysis \cite{farias,emin,eva} on the mobility of
small polarons also in conjunction with models
on polaronic high $T_c$ superconductivity
\cite{mott,mustre,fehske} have led to reconsider the
concept of self-trapped state which, although
being intimately related to the small size 
of the quasiparticle, {\it is not} synonymous
with small polaron state. Thus, if the narrowing of the polaron bandwidth (induced by
an increasing e-ph coupling) marks the
onset of the transition between large and
small states still there is a range of $g$
values for which the polaron, although spread over a few lattice sites only, is not trapped and retains mobility properties.
The self-trapping event can be instead associated
with a rapid but continuos effective mass increase which is precisely
located by looking at the curvature of the
logarithm of the effective mass versus $g$.
In our view this method, beyond embodying the full physical significance of the transition 
process, offers a simple criterion to select
a  "critical $g$ value" as
an inflection point either in the
logarithm of the effective mass or in its
first derivative. While the former case
would identify the point of most rapid 
increase of the effective mass, the occurence
of the latter case distinguishes a peculiar
point in the {\it mass versus g} plot although
the concavity-convexity change is absent.

In Fig.1(a), four plots of the polaron mass
(in units of the bare band mass) as a function of the $g$ coupling are reported
on while the corresponding curvatures of
the logarithm of the mass are shown in Fig.1(b). High optical phonon frequencies
are assumed. Our selected plots range from an
extreme adiabatic ($t=\,200meV$) to a moderately antiadiabatic ($t=\,10meV$) regime.
At a fixed $g$ the antiadiabatic polaron is
always heavier than the adiabatic polarons
but the peculiar points, resolved as the 
minima of the curves displayed in Fig.1(b),
lie at {\it decreasing} $g$ values by
decreasing the degree of adiabaticity.
Then, an antiadiabatic polaron may self-trap
already at $g \simeq 2.7$ weighing $m_{eff} \simeq 14$ while an extreme adiabatic polaron
self-traps only at $g \simeq 3.8$ with
$m_{eff} \simeq 123$. 
The correctness of our perturbative approach
is monitored by the $\lambda$ and $\alpha$
values which are larger than one in all the
displayed points consistently with the
assumption on the existence of small polarons.
In the intermediate adiabatic cases $t=\,100meV$ and $t=\,50meV$ we find
$m_{eff} \simeq 73$ at $g \simeq 3.5$ and
$m_{eff} \simeq 40$ at $g \simeq 3.3$ respectively. Thus, a 2D honeycomb net
seems to sustain mobile adiabatic polarons
in a window of strong coupling regimes 
approximately defined by $2.5 \le g \le 3.3$.
In Figs.2, the phonon frequencies are still
high, although  much
reduced with respect to Figs.1. As a main
effect the polaron masses are roughly doubled
while the
peculiar minima of the second derivative of
the logarithm of the effective mass (Fig.2(b)) 
do not shift substantially versus $g$ with respect
to the corresponding cases in Fig.1(b).
To emphasize the role of the lattice structure
we have compared the adiabatic polaron (with $t=\,100meV$)
in the present 2D honeycomb net with the previously
investigated \cite{io} square lattice. 
As an example, provided that:
i) the same overall e-ph coupling $g=\,3.38$ 
is taken,
ii) the same values of intra- and inter-molecular force
constants are assumed  ($\omega_0=\,50meV$
and $\omega_1=\,25meV$) and these values are consistent
with the strong coupling perturbative method,
we find $m_{eff} \simeq 140$ for the honeycomb net
polaron against $m_{eff} \simeq 1200$ for the square lattice.
Also the intermediate adiabatic polaron
($t=\,50meV$) behaves in a similar manner
being $m_{eff} \simeq 200$ in the honeycomb net
polaron against $m_{eff} \simeq 1700$ in the square lattice assuming the same input
parameters as above.
We have also considered the effect of the acoustical
branch of the phonon spectrum on the polaron properties.
In all cases, with different degree of adiabaticity, the minima of the second derivative (Fig.3(b)) occur
at much lower $g$ values than for optical polarons. However, no physical meaning can be attached to these values since they are well 
outside the range of validity of the Lang-Firsov 
based perturbative method. When the method holds
($g > 2.5$) no peculiar point can be resolved in
the polaron behavior versus $g$ which is anyway characterized by a huge mass renormalization.
Figs.(3) have been reported also to point out that the occurence of distinctive 
features in the mass or mass derivative
curves, far from being a cogent criterion for
the self-trapping event, just indicates a
trend which needs to be corroborated by the
analysis of other independent quantities.

\begin{figure}
\vspace*{12truecm}
\caption{(a)
Polaron masses (in units of bare band
electron mass) versus $g$ in adiabatic and 
antiadiabatic regimes with high frequency optical phonons: $\bar \omega=\,158meV$.
(b)
Second derivative of the logarithm
of the effective mass with respect to $g$.}
\end{figure}

\begin{figure}
\vspace*{12truecm}
\caption{(a)
Polaron masses (in units of bare band
electron mass) versus $g$ in adiabatic and 
antiadiabatic regimes. The characteristic frequency of the optical phonons is: $\bar \omega=\,79meV$.
(b)
Second derivative of the logarithm
of the effective mass with respect to $g$.}
\end{figure}

\begin{figure}
\vspace*{12truecm}
\caption{(a)
Polaron masses (in units of bare band
electron mass) versus $g$ in three adiabatic  regimes. The electron couples here to the
acoustical phonon branch.
(b)
Second derivative of the logarithm
of the effective mass with respect to $g$.}
\end{figure}

\section*{IV. Conclusions}

We have developed a perturbative study of the 
molecular crystal model assuming the existence
of strong electron-phonon coupling conditions
which favor the formation of small polarons.
The Lang-Firsov method permits to calculate the
mass renormalization for specific structures
once the phonon spectrum is known. Rather than applying the model to real systems
as previously done for simple lattices, 
we have examined whether alternative
structures as the 2D honeycomb net may host
polarons which are both small and mobile.
Infact we have found that adiabatic small polarons are  
lighter by roughly a factor eight than in the square
lattice once the same input parameters are assumed.
Adiabatic polarons have been studied as a function
of the e-ph coupling and distinctive points in the
mass behavior (versus $g$) have been selected and 
put in relation with a possible occurence of the
self-trapping event. Although not ultimative this
criterion seems plausible provided that such 
"self-trapping $g$ points" are obviously 
well within the range of
applicability of the Lang-Firsov scheme. We find
that small optical polarons in the
honeycomb net can self-trap if the electron-lattice
system couples in the range $3 < g < 4$. 
The exact
location of the transition depends on the adiabaticity
parameter with intermediate adiabatic polarons lying
in the lower portion of that range. Our results
suggest that adiabatic optical polarons can be 
mobile in the
honeycomb net although a strong coupling regime
holds. The small polaron effective mass is of order
$\simeq 5 - 50$ times the bare band mass before the
self-trapping point is attained, with more
adiabatic polarons being lighter once the {\it e-ph}
coupling is fixed.

This work has been done in Trequanda (Siena).

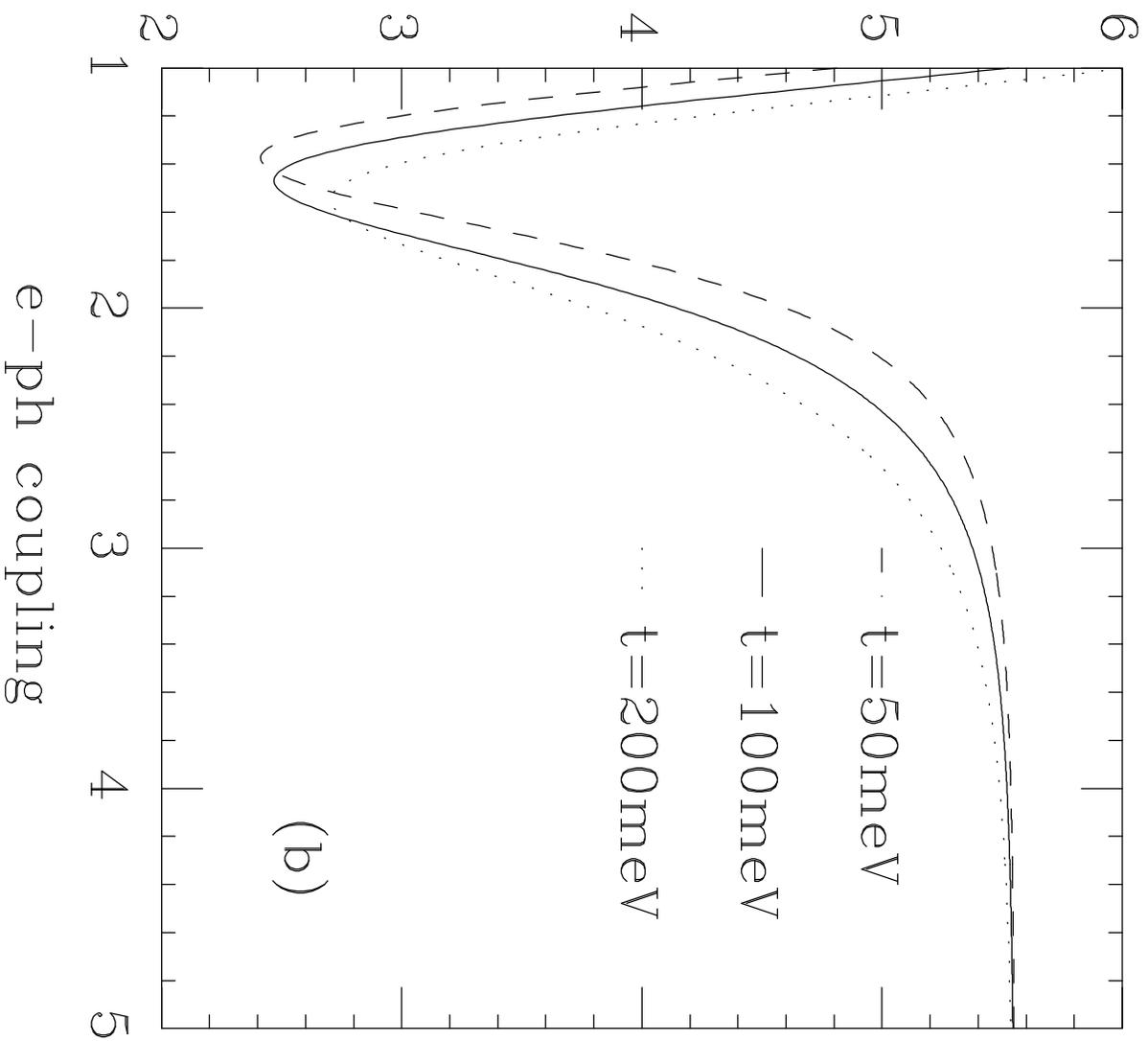
\end{document}